\documentstyle[12pt]{article}
\begin{document}
\begin{titlepage}
\begin{flushright}
TU-513 Revised\\
gr-qc/9611058
\end{flushright}
\ \\
\ \\
\ \\
\ \\
\begin{center}
{\LARGE \bf
A Comment on \\
 the Geometric Entropy \\
and  Conical Space
}
\end{center}
\ \\
\ \\
\begin{center}
\Large{
M.Hotta, T.Kato and K.Nagata
 }\\
{\it 
Department of Physics, Tohoku University,\\
Sendai 980-77, Japan
}\\
To appear in Classical and Quantum Gravity
\end{center}
\ \\
\ \\

\begin{abstract}
It has been recently pointed out that 
 a definition of the geometric entropy 
 using the partition function in a conical space   
 does not in general lead to a positive definite quantity.
For a scalar field model with a non-minimal coupling 
 we clarify the origin of the anomalous behavior from the viewpoint 
of the canonical formulation.
\end{abstract}

\end{titlepage}

\section{
Introduction
}

\ \\

The concept of the geometric entropy  
 has attracted much attention 
 for the past several years in elucidating the black hole entropy
 \cite{BKLS}-\cite{LW2}.
 In this investigation a possibility was pointed out that 
the ultraviolet divergence of the entropy 
may exactly be identified with the divergence
 of the gravitational constant in the perturbative treatment of 
 quantum gravity \cite{SU}.  
However  
the hypothesis on the equivalence of the two divergences 
has a crucial weak point in its discussion \cite{SU}-\cite{DLM}. 
For a scalar field coupled non-minimally to the curvature
 like ${\cal L}_{\xi} = -\xi R\phi^2 /2$ with $\xi >1/6$
 and a gauge vector field in lower than eight dimensions,
 divergent corrections of the gravitational coupling take  
 negative values. On the other hand, the divergence of 
 the geometric entropy is expected to be positive. Thus 
 the two quantities does not seem to be identical.

Even though such a severe problem is exposed, Larsen and Wilczek   
 have recently argued \cite{LW} that the two divergences 
of the geometric entropy
 and of the gravitational constant, nevertheless,  
coincide   for the case of fields of spin 0, 1/2 and 1  
 if one specifies the definition of the entropy 
by use of a partition function in a conical space. 
 For example, their definition   
 works well for a minimally coupled 
scalar and  a spinor field   such that it gives 
 positive values precisely equal to the correction 
to the gravitational constant.

Their idea is quite appealing, but we think that 
there still remains an ambiguous point in their argument. 
 This is 
because their entropy still yields  
 negative values in the previously mentioned cases of    
the non-minimally coupled field and the gauge vector field. 
 In general a definition of the entropy requires   
that the entropy  
is ensured to be positive and possesses a certain statistical origin.
 However they, instead, cite a work \cite{CW} 
in which it is only proved that the entropy of a minimally coupled
 field can be computed rigorously using the partition function on a cone.

In the present work we analyze carefully 
their definition for the non-minimal coupling case and
 argue that it cannot be interpreted directly 
as the geometric entropy.

The contents in subsequent sections are as follows.
In Section 2 we briefly review definition of the geometric 
entropy using the partition function in a conical space 
taking a minimally coupled scalar field. 
 In Section 3 
 we investigate a proposal given by Larsen and Wilczek to
define the geometric entropy for a non-minimally coupled field.
 In Section 4 we show that besides 
the standard entropy term: $ -Tr \rho \ln \rho$ 
their definition includes a correction term 
which has no statistical origin, thus this is unacceptable. 
This conclusion is similar to 
the result for a gauge vector field  obtained by Kabat \cite{K}.

\section{
Geomeric Entropy, Replica Trick 
and Partition Function on the Cone
}

\ \\

In this section we review briefly a recent advance 
\cite{CW}, 
that is, 
how the geometric entropy can be defined using the replica method 
and the Euclidean path integral 
taking an example of the minimally coupled scalar field in D dimensions.
\\

To investigate the divergent part of the 
 entropy, it is sufficient to discuss only the minimally coupled
 massless case. 
 The action takes the form of 
$$
S_{min} = -\frac{1}{2} \int d^D x  \sqrt{-g_L} g^{\mu\nu}_L 
\frac{\partial\phi}{\partial x^\mu }
\frac{\partial\phi}{\partial x^\nu }
$$
where the form of the  flat metric  $g_{L\mu\nu}$ 
 can be taken in a variety of ways. 
In this discussion we use only 
two forms of the metric:
\begin{eqnarray}
ds^2 = -(dx^0) ^2 + (dx^1 )^2 + \sum^{D-1}_{i=2} (dx^i )^2 \label{01}
\end{eqnarray}
and
\begin{eqnarray}
ds^2 = -r^2 dt^2 + dr^2 + \sum^{D-1}_{i=2} (dx^i )^2.\label{02}
\end{eqnarray}
The latter form, eqn(\ref{02}), is usually called Rindler metric. 
The Rindler metric has a timelike Killing vector 
 $\partial_t$. Thus a conserved 
Hamiltonian $H_R$ exists due to the symmetry:
$$
H_R = 
\int^\infty_0 dr r
 \left[ \frac{1}{2} \Pi^2 +
\frac{1}{2}
\left( \frac{\partial \phi}{\partial r} \right)^2
+\frac{1}{2}
\sum^{D-1}_{i=2} \left( \frac{\partial \phi}{\partial x^i} \right)^2
\right]
$$
where $\Pi$ is a conjugate momentum of $\phi$.

Let us first start from the standard definition of the entropy.
Consider the wavefunctional of the Minkowskian vacuum state
in the cartesian coordinates, eqn(\ref{01}):
$$
\langle \phi |0 \rangle =\Psi_o [\phi(x^1,\cdots ,x^{D-1})] 
=\Psi_o [\phi_+ , \phi_- ],
$$
where 
$\phi_+ = \phi (x^1 \geq 0 )$ and $\phi_- =\phi (x^1 <0 )$. 
Tracing over fields in the $x^1 <0$ half-space 
 reduces the pure wavefunctional $\Psi_o$ to a mixed 
density matrix:  
\begin{eqnarray}
&&\langle \phi_{1+} |\rho |\phi_{2+} \rangle=
\rho [\phi_{1+}  ,
\phi_{2+}  ]\nonumber\\
&=& \int {\it D} \phi_- 
\Psi_o [\phi_{1+} , \phi_- ]
\Psi^\ast_o [\phi_{2+} , \phi_- ].\label{003}
\end{eqnarray}
The definition of geometric entropy is given as follows.
\begin{eqnarray}
S_{geo} \equiv -Tr\rho\ln\rho .\label{1}
\end{eqnarray}

It is  worthwhile       
 to recall the well known fact
 \cite{US,KS}  
 that the density matrix (\ref{003}) takes a thermal form such as
\begin{eqnarray}
\rho = \frac{e^{-2\pi H_R } }{Tr(e^{-2\pi H_R })} .\label{0}
\end{eqnarray}
This relation  will be used in the following analysis.

The definition (\ref{1}) can be also rewritten  
by use of the replica trick:
\begin{eqnarray}
S_{geo} = \left( 1-n \frac{\partial}{\partial n} \right)
\ln Tr  \rho^n |_{n=1} .\label{2}
\end{eqnarray}
This expression proposes a nice way of defining the geometric 
entropy invoking the thermalization theorem, eqn(\ref{0}),
 and path-integral formulation.
Substituting eqn(\ref{0}) into eqn(\ref{2}), we obtain
\begin{eqnarray}
S_{geo} = \left( 1-n \frac{\partial}{\partial n} \right)
\ln Tr \left[(e^{-2\pi  H_R })^n \right] |_{n=1} ,
\end{eqnarray}
Here notice that a factor $(Tr e^{-2\pi  H_R} )^{-n}$ coming from
 normalization of the density matrix does not contribute
to the entropy itself.
Using the standard path integration, it is proved that
 the kernel part can be regarded as a partition function
of the field in a periodic manifold in time.  
\begin{eqnarray}
&&Tr\left[ \left( e^{-2\pi  H_R } \right)^n \right]=Tr e^{-2\pi n H_R }
\nonumber\\ 
&=&
\int_{periodic} D\Pi D\phi 
\exp
\left[
i\int^{2\pi n}_0 d\tau \int^\infty_0 dr \int d^{D-2} x \Pi 
\frac{\partial \phi}{\partial \tau} \right.\nonumber\\
&&\left. -\int^{2\pi n}_0 d\tau
\int^\infty_0 dr r \int d^{D-2} x
\left[ \frac{1}{2} \Pi^2 +
\frac{1}{2}
\left( \frac{\partial \phi}{\partial r} \right)^2
+
\frac{1}{2}\sum^{D-1}_{i=2} 
\left( \frac{\partial \phi}{\partial x^i} 
\right)^2
\right]
\right]\nonumber\\
&=&
\int_{periodic} D\phi 
\exp \left[
-\int^{2\pi n}_0 d\tau
\int^\infty_0 dr   \int d^{D-2} x
\left[ \frac{1}{2r } 
\left( \frac{\partial \phi}{\partial \tau} \right)^2
+
\frac{r}{2}
\left( \frac{\partial \phi}{\partial r} \right)^2
+
\frac{r}{2}\sum^{D-1}_{i=2} 
\left( \frac{\partial \phi}{\partial x^i} 
\right)^2
\right]
\right]
\nonumber
\end{eqnarray}
where a periodic condition, $\phi(\tau + 2\pi n) =\phi (\tau)$,
 is imposed in the path integral. 
By virtue of a term:
$$
 \frac{1}{2r } 
\left( \frac{\partial \phi}{\partial \tau} \right)^2
$$
in the path-integral action, configurations with 
$\partial_{\tau} \phi (r=0) \neq 0$ are highly suppressed
 and thus an equation:
$$
\phi(\tau, r =0) =\phi (0,0)
$$
can be used in the above path integral. 
Therefore it is justified that
 the surface region defined by $r=0$ is exactly a point, that is,
 the space has a conical structure around $r=0$.  
Consequently we obtain an amazing expression of the geometric
entropy \cite{CW}:
\begin{eqnarray}
S_{geo}
 = \left( 1-n \frac{\partial}{\partial n} \right)
\ln Z(n) \left|_{n=1} \right., \label{5}
\end{eqnarray}
where $Z(n)$ is a partition function of the scalar field 
in a conical space of the deficit angle $\delta=2\pi (1-n) $:
$$
Z(n)
=
\int_{cone} {\it D}\phi
  \exp \left[ -\frac{1}{2} \int^{2\pi n}_0 d\tau
\int^\infty_0 dr
\int_{R^{D-2}} d^{D-2} x \sqrt{g}
  g^{\mu\nu} \partial_\mu \phi \partial_\nu \phi
 \right]
$$
and
\begin{eqnarray}
&&g_{\mu\nu} dx^\mu dx^\nu 
=
r^2 d\tau^2 +dr^2 + \sum^{D-1}_{a=2} (dx^a )^2 .\nonumber\\
&&(0\leq \tau \leq 2\pi n,\ 0\leq r \leq \infty.)\nonumber
\end{eqnarray}

Due to the ultraviolet behavior of the partition function $Z(n)$
 the geometric entropy calculated from eqn(\ref{5}) 
 is also divergent. We thus need some regularization scheme and 
adopt  the heat kernel regularization:
\begin{eqnarray}
\ln Z(n) =
-\frac{1}{2} \ln Det(-\nabla^2 )
\rightarrow\frac{1}{2} \int^{\infty}_{\epsilon^2} \frac{dT}{T}
Tr e^{-T(-\nabla^2 )},
\end{eqnarray}
where $\epsilon$ is a covariant short distance cutoff.
Then the most singular term of the 
geometric entropy is evaluated straightforwardly 
   as follows \cite{CW,K,LW}:
\begin{eqnarray}
S_{geo} \sim \frac{A^{(D-2)}_{\perp} }{4} 
\frac{2}{3(D-2)} \left[\frac{1}{4\pi\epsilon^2} \right]^{D/2 -1},
\label{77}
\end{eqnarray}
where $A^{(D-2)}_{\perp} = \int \prod^{D-1}_{i=2} d x^i$.

The original definition of the entropy: 
$-Tr \rho\ln\rho $ is  known  
 to take a non-negative value. Consistent
 with this result,  
the positive term appears in eqn(\ref{77}).

Furthermore it has been established \cite{K,LW} 
that this divergence is automatically  
renormalized by the gravitational constant $G$  
if the divergent term is added
to the bare Bekenstein-Hawking entropy, 
$A^{(D-2)}_{\perp} /(4G_o )$:
$$
\frac{1}{G} = \frac{1}{G_o } +
\frac{2}{3(D-2)} \left[\frac{1}{4\pi\epsilon^2} \right]^{D/2 -1} .
$$
This equivalence between the entropy and the gravitational 
constant renormalizations  is proved explicitly 
 not only for the minimally coupled scalar field but also
 for a spinor field \cite{LW2,K,LW} 
 in any spacetime dimension by use of a similar definition to 
eqn(\ref{5}) .

Thus it can be summarized that 
the definition (\ref{5}) works effectively
 for the minimally coupled field and the spinor field.

\section{
Non-Minimal Coupling Case 
}

\ \\
 
Larsen and Wilczek \cite{LW} argue  
 that a proper definition of the geometric entropy 
is given by eqn(\ref{5})   
 not only for 
the minimally coupled scalar field and the spinor field
 but also for other fields such as 
a non-minimally coupled scalar field and 
 a gauge vector field by substituting  
 each partition function of the field on the cone
 into eqn(\ref{5}).

In this and the next section we examine  
 the non-minimal coupling case and check 
 whether this definition
  is truly suitable for the geometric 
entropy or not.

Let us start from considering a non-minimally 
 coupled massless field
 in a general static space.
The Lorentzian action reads as
$$
S_{nm} = \int d^D x \sqrt{-g_{L}} 
\left(
-\frac{1}{2} g^{\mu\nu}_{L} 
\partial_\mu \phi 
\partial_\nu \phi 
-\frac{\xi}{2} R_{L} \phi^2
\right),
$$
where $g_{L\mu\nu}$ is given by 
\begin{eqnarray}
g_{L\mu\nu} dx^\mu dx^\nu
= -N(\vec{x} )^2 (dx^0 )^2 +h_{ab} (\vec{x})dx^a dx^b,
\end{eqnarray}
with $N$ and $h_{ab}$ independent of $x^0$.

Now let us define an entropy of the field
 in a similar way to that \cite{LW} :
\begin{eqnarray}
S_{\xi}
 = \left( 1-n \frac{\partial}{\partial n} \right)
\ln Z(n, \xi) \left|_{n=1} \right., \label{6}
\end{eqnarray}
where 
$$
Z(n ,\xi)
=
\int_{periodic} {\it D}\phi
  \exp \left[ -\frac{1}{2} \int^{2\pi n}_0 d\tau 
\int d^{D-1} x \sqrt{g}\left(
  g^{\mu\nu} \partial_\mu \phi \partial_\nu \phi
+\xi R \phi^2\right)
 \right], 
$$
\begin{eqnarray}
g_{\mu\nu} dx^\mu dx^\nu
= N(\vec{x} )^2 d\tau^2 +h_{ab} (\vec{x})dx^a dx^b,
\ (0\leq \tau \leq 2\pi n) ,\label{7} 
\end{eqnarray}
and the periodic condition, $\phi(\tau + 2\pi n) =\phi (\tau)$, 
 is imposed. 
Then we can prove formally the positivity of the entropy $S_\xi$. 
 To show this, rewrite the partition function $Z(n,\xi)$
 as follows. 
\begin{eqnarray}
&&Z(n,\xi)\nonumber\\
&=&
\int_{periodic}  D\phi\exp 
\left[
-\int^{2\pi n}_0 d\tau \int d^{D-1} x N\sqrt{h}
\left[
\frac{1}{2N^2} \dot{\phi}^2 
+\frac{1}{2} h^{ab}
\partial_a \phi \partial_b \phi
+\frac{\xi}{2} R_{(L)} \phi^2
\right]
\right]\nonumber\\
&=&\int_{periodic} D\Pi D\phi
\exp
\left[
i\int^{2\pi n}_0 d\tau \int d^{D-1} x \Pi \dot{\phi} \right.\nonumber\\
 &-&\left.
\int^{2\pi n}_0 d\tau 
\int d^{D-1} x N\sqrt{h}
\left[
\frac{1}{2}
 \left(
\frac{\Pi}{\sqrt{h}}
\right)^2
+\frac{1}{2}h^{ab} \partial_a \phi \partial_b \phi
+\frac{\xi}{2} R \phi^2
\right]
\right]\nonumber\\
&=&
Tr e^{-2\pi n \hat{H}_{can} }
\end{eqnarray}
where 
$$
\hat{H}_{can} 
=
\int d^{D-1} x N\sqrt{h}
\left[
\frac{1}{2}
 \left(
\frac{\hat{\Pi}}{\sqrt{h}}
\right)^2
+\frac{1}{2}h^{ab} \partial_a \hat{\phi} \partial_b \hat{\phi}
+\frac{\xi}{2} R \hat{\phi}^2
\right]
$$
and $\hat{\Pi}$ is a canonical momentum operator conjugate to 
$\hat{\phi}$.
Substituting $Z(n,\xi)=Tr e^{-2\pi n \hat{H}_{can}} $
into eqn(\ref{6}) and  using a {\it "fact"} that 
$\hat{H}_{can}$ is independent of $n$ , 
we get 
\begin{eqnarray}
S_{\xi} &=&\ln Tr e^{-2\pi \hat{H}_{can}} 
+2\pi  \frac{Tr( \hat{H}_{can} e^{-2\pi  \hat{H}_{can}} ) }
{Tr e^{-2\pi  \hat{H}_{can}} }
\nonumber\\
&=& -Tr \varrho \ln \varrho,\nonumber
\end{eqnarray}
where
$$
\varrho = \frac{e^{-2\pi  \hat{H}_{can}} }
{Tr e^{-2\pi  \hat{H}_{can}} } .
$$
Therefore if the analysis is entirely correct,
the entropy $S_{\xi}$  really has  
 an explicit statistical origin and is guaranteed  to yield 
a non-negative value.

It should be stressed here that 
the metric of the cone, which is of our concern, is also written
in the above form (\ref{7}) with $N =x^1 =r$ and
 $h_{ab} =\delta_{ab}$. Moreover, 
from the replica trick point of view as seen in Section 2,
 the entropy $S_{\xi}$ for the conical space
could be 
naively identified with the geometric entropy $S_{geo}$ itself. 
Larsen and Wilczek \cite{LW} 
 insist that  eqn(\ref{6}) with the conical metric 
 defines  naturally the geometric entropy
 even for the non-minimally coupled scalar field  and 
calculate it explicitly.
Then they give the following result: 
\begin{eqnarray}
S_{geo} =S_{\xi} \sim \frac{A^{(D-2)}_{\perp} }{4} 
\frac{4}{(D-2)}\left(\frac{1}{6} -\xi \right)
 \left[\frac{1}{4\pi\epsilon^2} \right]^{D/2 -1},\label{15}
\end{eqnarray}
 adopting the heat kernel regularization just like
 in  Section 2.

Despite the apparent naturalness of the definition, 
 it is noticed that
 $S_\xi$ fails to take a non-negative value for $\xi >1/6$. 
Though this negativeness of the entropy seems very queer,
 they still argues that the above result is natural and correct 
\cite{LW}.

Our opinion for the negative value  
is quite different from that of Larsen and Wilczek
 and rather similar to that of Kabat \cite{K}.  
 It should be emphasized, we believe,  
that any definition of the geometric entropy must 
both intrinsically be non-negative  and 
possess a manifest statistical meaning. 
The negative value just makes us doubt deeply of the validity
 of their definition.

In fact, as to spaces with conical structure,
 the previous argument on positivity of $S_\xi$ 
 is too formal and 
clearly incorrect . 
It completely misses out an effect of 
 the delta functional curvature at $r=0$.  
We shall argue in the next section, 
treating the conical singularity carefully,  
that the quantity $S_\xi$ in the conical space 
cannot be identified exactly with the geometric entropy $S_{geo}$.

\section{
Validity of a Definition of the Geometric Entropy
using Partition Function on the Cone
}

\ \\

The definition of the geometric entropy proposed by 
Larsen and Wilczek \cite{LW} is certainly interesting. However
 in this section we argue that their definition needs  
 some modification for that to be regarded as 
the entropy of the non-minimally coupled scalar field.

In order to properly treat the conical singularity,
 let us first express the conical space as a limit
of a non-singular static curved space.
Consider the following metric:
$$
ds^2 =\sum_{n,m =0}^1 \tilde{g}_{nm} dx^n dx^m
+ 
\sum^{D-1}_{i=2} (dx^i )^2=
F(r)^2 d\tau^2 +dr^2 + 
\sum^{D-1}_{i=2} (dx^i )^2 
$$
with $0\leq \tau \leq 2\pi n$ , 
$ F(r=0) =0$, $F(r\sim\infty )=r$ and $\partial_r F(r=0) =c$.
Here the point $(\tau=2\pi n,r, x^i )$ is identified with $(0,r,x^i )$.

The scalar curvature $R$ is obtained by a simple calculation:
$$
R=4\pi(1-c n)
 \frac{1}{\sqrt{\tilde{g}} } \delta (x^0 ) \delta (x^1 ) 
-\frac{2}{F}\frac{\partial^2 F}{\partial r^2}.
$$
To remove the delta function in the curvature, we set 
$$
c=\frac{1}{n} .
$$
Note that this choice of $c$ demands a compensation 
that the metric itself has $n$ dependence. 
This $n$ dependence is crucial to understand the anomalous 
negative value of $S_\xi$, as shown subsequently.
To make our analysis more concrete,
 let us give an explicit example form of $F$ such as 
\begin{eqnarray}
F(r,n) = r\left[\frac{1}{n} +\left( 1-\frac{1}{n} \right) 
\frac{\lambda^2 r^2}{1+\lambda^2 r^2} \right]. \label{8}
\end{eqnarray}
If we take $\lambda\rightarrow \infty$ limit,
the conically flat space  appears again:
$$
\lim_{\lambda \rightarrow \infty} F(r,n) = r.
$$
Eqn(\ref{8}) yields 
a scalar curvature without the delta function:
\begin{eqnarray}
R(r,n)=
4\lambda^2 \frac{3-\lambda^2 r^2}{(1+\lambda^2 r^2 )^2} 
\frac{1-n}{1+n\lambda^2 r^2} .
\label{9}
\end{eqnarray}
For this metric it is easy to repeat the same argument in Section 3
and we obtain
\begin{eqnarray}
&&Z(n ,\xi)= Tr e^{-2\pi n \hat{H}_{can} (n) }\nonumber\\
&=&
Tr \exp \left[
-2\pi n \int^\infty_0 dr F(r,n) \int d^{D-2} x 
\left[
\frac{1}{2} \hat{\Pi}^2
+\frac{1}{2}\delta^{ab} \partial_a \hat{\phi} \partial_b \hat{\phi}
+\frac{\xi}{2} R(r,n) \hat{\phi}^2
\right] 
\right].\nonumber
\end{eqnarray}
It is worth noting  that the $n$ dependence still remains in 
$\hat{H}_{can} (n)$. Thus when  the derivative with respect
 to $n$ in eqn(\ref{6}) is taken, we cannot neglect naively 
a term proportional to 
$\partial_n \hat{H}_{can}$:
\begin{eqnarray}
S_{\xi} &=& \left( 1-n\frac{\partial}{\partial n} \right) 
\ln Tr e^{-2\pi n \hat{H}_{can} } |_{n=1}
\nonumber\\
&=&
-Tr \varrho \ln \varrho
+
2\pi Tr \left[ \varrho
\frac{\partial \hat{H}_{can}}{\partial n}|_{n=1}
  \right] ,
\end{eqnarray}
where
$$
\varrho = \frac{e^{-2\pi  \hat{H}_{can} (1)} }
{Tr e^{-2\pi  \hat{H}_{can}(1)} } .
$$
The first term $-Tr \varrho \ln \varrho$ 
 clearly takes  a non-negative value and should be identified
 with the geometric entropy $S_{geo}$ 
after taking $\lambda\rightarrow\infty$. 
Now a crucial question is whether the limit of the second 
correction term:
$$
\lim_{\lambda\rightarrow \infty}
2\pi Tr \left[ \varrho\frac{\partial \hat{H}_{can}}{\partial n}|_{n=1}
  \right]
$$ 
vanishes or not. It is easily shown from eqns(\ref{8}) and 
(\ref{9}) that the following relations are satisfied: 
\begin{eqnarray}
&&\lim_{\lambda\rightarrow\infty}
\frac{\partial F}{\partial n} |_{n=1} =0,\nonumber\\
&&\lim_{\lambda\rightarrow\infty}
\frac{\partial R}{\partial n} |_{n=1} = -2\frac{1}{r} \delta(r).\nonumber
\end{eqnarray}
Using these relations  we finally obtain the following result with the 
non-vanishing correction term:
\begin{eqnarray}
\lim_{\lambda\rightarrow \infty} S_{\xi}
&=&
-Tr \rho \ln \rho
-
2\pi\xi  Tr 
\left[ \int d^{D-2} x \phi(r=0, x^2, \cdots,x^{D-1} )^2
 \rho \right]\nonumber\\
&=&
-Tr \rho \ln \rho
-
2\pi\xi A^{(D-2)}_{\perp}
\frac{\langle 0_M| \phi(0)^2 |0_M \rangle}
{\langle 0_M |0_M \rangle}, \label{12}
\end{eqnarray}
where $$\rho = \frac{e^{-2\pi H_R } }{Tr(e^{-2\pi H_R })}, $$
$|0_M \rangle $is the Minkowskian vacuum state and 
 we have used  the thermalization theorem (\ref{0}):
$$
Tr_{x^1 <0} \left[ |0_M\rangle\langle 0_M |\right] = \rho.
$$
We also comment that even when 
the $\lambda\rightarrow\infty$ limit
 is performed before taking $n\rightarrow 1$ 
the same result (\ref{12}) appears from relations
for the conical space:
$$
R=4\pi(1-n)
 \frac{1}{\sqrt{\tilde{g}} } \delta (x^0 ) \delta (x^1 ) ,
$$
 and 
$$
\lim_{\lambda \rightarrow \infty} Z(n,\xi)
=
Tr\exp \left[
-2\pi \xi \int d^{D-2} x \phi(r=0)^2
-2\pi n H_R (\xi)
\right],
$$
where
$$
H_R (\xi) = H_R -\xi \int d^{D-2} x \phi(r=0)^2 .
$$

Consequently the entropy 
 proposed by Larsen and Wilczek
 evidently differs from the geometric entorpy $S_{geo}$ due to
the existence of the non-statistical correction term proportional to 
$\langle \phi (0)^2\rangle$ when $\xi \neq 0$. 
Because $\langle \phi (0)^2\rangle$ is positive,  
 irrespective of the detail of the regularization, 
 the value of $S_{\xi}$ reverses its sign for positive $\xi$ large enough. 
The geometric entropy for the non-minimal coupling case 
is more naturally  defined by $S_\xi$ in eqn(\ref{6}) 
with the conical metric {\it and} 
$\xi =0$, just like the minimal coupling case.

It has been already 
pointed out for the gauge vector field case by Kabat \cite{K} 
that the similar entropy correction term like in eqn(\ref{12}) 
follows and that the entropy defined by use of
 the partition function 
cannot be equated with the geometric entropy.

We can also prove explicitly that 
the $<\phi(0)^2>$ term in eqn(\ref{12}) is precisely 
equal to the deviation of $S_\xi$  in eqn (\ref{15}) from 
the geometric entropy, using the heat kernel regulator. 
The proof is as follows.
\begin{eqnarray}
\Delta S&=&-2\pi\xi A^{(D-2)}_{\perp} 
\frac{<0_M | \phi^2 (0,0) |0_M >}{<0_M | 0_M >}\nonumber\\
&=&
-2\pi \xi\frac{A^{(D-2)}_{\perp}}{V_D } \int d^D x G_E (x,x) \nonumber\\
&=&
-2\pi \xi \frac{A^{(D-2)}_{\perp} }{V_D }
Tr\left[\frac{1}{-\partial^2 } \right]\nonumber\\
&\rightarrow&
-2\pi\xi\frac{A^{(D-2)}_{\perp} }{V_D }
\int^\infty_{\epsilon^2} dT\  Tr e^{T \partial^2} \nonumber\\
&=&
-2\pi\xi \int^\infty_{\epsilon^2} dT \frac{A^{(D-2)}_{\perp}}
{(4\pi T)^{D/2}  }
\sim
-\xi\frac{A^{(D-2)}_{\perp} }{4} \frac{4}{D-2}
\left[\frac{1}{4\pi \epsilon^2 } \right]^{D/2-1}\nonumber
\end{eqnarray}
where $V_D = \int  d^D x$ and 
the following relations have been  used:
\begin{eqnarray}
G_E (x, y) &=&<\phi(x) \phi(y) >\nonumber\\
&=&\frac{1}{Z_o}
\int D\phi \phi(x) \phi (y)
\exp\left[
-\int d^4 x \frac{1}{2} (\partial \phi)^2 \right] \nonumber\\
&=&
\frac{\delta}{\delta J(x) }
\frac{\delta }{\delta J(y) }
\frac{1}{Z_o}\int D\phi \exp\left[
-\int d^4 x \left[
\frac{1}{2} (\partial \phi)^2
+J\phi
\right]
 \right] |_{J=0} \nonumber\\
&=& 
-\frac{1}{\partial^2 } \delta^4 (x-y) .\nonumber
\end{eqnarray}
Thus the correction term $\Delta S$ precisely reproduces 
the term linearly depending on $\xi$ in eqn(\ref{15}), 
which demonstrates that 
the quantity (\ref{6}) is negative. 
\ \\
\ \\
\ \\
{\bf Acknowledgement}\\

We would like to thank M.Yoshimura for reading the manuscript and
 giving us helpful comments.
\ \\
\ \\
\ \\
{\it Note added.}\\

After submitting this paper we received a few comments.

We were informed that 
the negative entropy for the non-minimal coupling case has 
independently been derived by S.N.Solodukhin\cite{S}  
using the heat kernel method.

Another comment is that the relation in eqn(\ref{12}) 
 has been also derived in the different context, namely, in 
a special class of the induced gravity theory by 
V.P.Frolov, D.V.Fursaev and A.I.Zelnikov \cite{FFZ}.

\end{document}